\documentclass[12pt]{iopart}
\usepackage{graphicx}
\usepackage{bm}
\newcommand{\beq}{\begin{equation}}
\newcommand{\eeq}{\end{equation}}
\newcommand{\beqa}{\begin{eqnarray}}
\newcommand{\eeqa}{\end{eqnarray}}

\newcommand{\rhov}{\mbox{\boldmath$\rho$}}

\begin{document}
\title{Frictionless dynamics of Bose-Einstein condensates 
under fast trap variations}

\author{J. G. Muga$^{1}$, Xi Chen$^{1,2}$, 
A. Ruschhaupt$^{3}$, 
D. Gu\'ery-Odelin$^{6}$}

\address{$^{1}$ Departamento de Qu\'{\i}mica-F\'{\i}sica,
UPV-EHU, Apdo 644, 48080 Bilbao, Spain}

\address{$^{2}$ Department of Physics, Shanghai University,
200444 Shanghai, P. R. China}

\address{$^{3}$ Institut f\"ur Theoretische Physik, Leibniz
Universit\"{a}t Hannover, Appelstra$\beta$e 2, 30167 Hannover,
Germany}

 

\address{$^{6}$Laboratoire Collisions Agr\'egats R\'eactivit\'e, CNRS UMR 5589, IRSAMC, Universit\'e Paul Sabatier, 118 Route de Narbonne, 31062 Toulouse CEDEX 4, France}

\begin{abstract}
A method is proposed to design the time dependence of the trap frequency  
and achieve in a short time 
an adiabatic-like (frictionless) evolution of Bose-Einstein condensates
governed by the Gross-Pitaevskii equation. Different cases depending on
the effective dimension of the trap and the interaction regimes are considered. 2D  traps are particularly suitable as the method can be applied without the need to impose any additional 
time-dependent change in the strength of the interatomic interaction or 
a Thomas-Fermi regime as it occurs for 1D and 3D traps.             
\end{abstract} 



\section{Introduction}
In order to manipulate 
Bose-Einstein condensates for different applications it is important to 
study and control their response to time-dependent changes of the confining fields. A natural approach to avoid undesired excitations
is to modify the trap adiabatically, i.e., very slowly, so that,
if the initial state is in the ground state 
the final state will be the ground state as well.
However, this may require very long times and become impractical. 
Faster changes are thus a desirable objective
but they will in general induce excitations and
oscillations (inner frictional heating \cite{Ronnie}),
so that the proportion of the ground state in the final state may be small \cite{Shlyap96,cd,Uhlmann}. These
difficulties  raise the question addressed in this paper:  Is it possible to change the trap in a very short time, taking the condensate, up to a global phase, to the same state that would be reached after a slow (adiabatic) process? This question has been answered recently in the affirmative for cooling expansions within the framework of the linear Schr\"odinger equation \cite{xi2}. For preliminary work see \cite{Schmidt,root2}. 
The method used to design the time-dependence of the trap frequency was based on  
Lewis-Riesenfeld invariants of motion \cite{LR69}
and simple inverse scattering techniques that had been applied for complex potential optimization \cite{Palao}. Our objective here is to analyze if and how the same techniques used in that simple case can be adapted to non-linear interactions and systems described by a Gross-Pitaevskii (GP) equation. As we shall 
see, the applicability of the method will depend critically on the effective dimension of the trap.  
We shall first discuss for simplicity with some detail
one dimensional (1D) traps, and then 2D and 3D traps subjected to time-dependent frequencies. By 1D traps we mean quasi-1D
cigar-shaped traps with tight (fixed) transversal confinement where the axial frequency is varied in time; similarly 2D traps 
are quasi-2D disk-shaped traps with tight, fixed, axial confinement
in which the transversal frequency is varied; and finally, the 
3D traps refer to harmonic traps with spherical symmetry. 
We assume in all cases that a GP equation can be derived corresponding to each dimensionality, and use 
$g$ generically for the coupling parameter of the non-linear term even though it  
is different for the three cases \cite{Salas}. 
\section{One dimensional traps\label{odt}}
Our starting point is the effective 1D Gross Pitaevski equation 
for the longitudinal ($x$) direction in an elongated cigar trap,
\beq
\label{gpe}
i\hbar \frac{\partial \psi}{\partial t}=\left[-\frac{\hbar^2}{2m}
\frac{\partial^2}{\partial x^2}+\frac{1}{2}m\omega(t)^2x^2+g|\psi|^2\right]\psi,
\eeq
$g$ being the coupling parameter.
The application of the invariant concept here is not as simple 
as for the Schr\"odinger equation \cite{invGP},
so we shall use instead an approach which leads in that 
case to the same results. The idea is to assume    
for the wavefunction the ansatz \cite{Cas} 
\beq\label{ans}
\psi(x,t)=e^{-\beta(t)}e^{-\alpha(t) x^2}\phi(x,t). 
\eeq
Substituting this into Eq. (\ref{gpe}), and using the scaling
$\rho=x/b$ and redefined wavefunction $\Phi(\rho,t)=\phi(x,t)$, 
we get 
\beqa
&&i\hbar\frac{\partial \Phi}{\partial t}=-\frac{\hbar^2}{2m}\frac{1}{b^2}
\frac{\partial^2 \Phi}{\partial \rho^2}
+\left[\frac{1}{2}m\omega(t)^2+i\hbar\dot{\alpha}
-\frac{2\hbar^2}{m}\alpha^2\right]
b^2\rho^2\Phi
\nonumber\\
&+&
\left[ge^{-(\alpha+\alpha^*)x^2}e^{-(\beta+\beta^*)}|\Phi|^2\right]\Phi
+\left[i\hbar\dot{\beta}+\frac{\hbar^2\alpha}{m}\right]\Phi
+\left[2\frac{\hbar\alpha}{m}+i\frac{\dot{b}}{b}\right]\hbar\rho\frac{\partial \Phi}{\partial\rho},
\label{bigeq}
\eeqa
where the dot means derivative with respect to time. 
Let us now impose that the coefficients in square brackets [...] of the last two terms  
vanish. This means that (we assume $b$ real) 
\beq
\beta=\frac{1}{2}\ln b,\;\;\;\;
\alpha=-\frac{im}{2\hbar}\frac{\dot{b}}{b},
\label{alpha}
\eeq
and  
%
$e^{-(\alpha+\alpha^*)x^2}e^{-(\beta+\beta^*)}= b^{-1}.$
%
Suppose now that the coefficient of $b^2 \rho^2\Phi$ in    
(\ref{bigeq}) is made constant, equal to $m\omega_0^2/(2b^4)$
(for an alternative see the final discussion), where $\omega_0=\omega(0)$.  
Using (\ref{alpha}) this is equivalent to imposing
for $b$ and $\omega(t)$ an Ermakov equation,
\beq
\ddot{b}+\omega(t)^2 b=\frac{\omega_0^2}{b^3}.
\label{erma}
\eeq
It is useful to express the resulting wave equation 
in terms of a new scaled time,
\beq
\label{tau2}
\tau(t)=\int_0^t \frac{dt'}{b^2},
\eeq
and wavefunction $\Psi(\rho,\tau)=\phi(\rho,t)$, 
\beq
\label{td2}
i\hbar\frac{\partial \Psi}{\partial \tau}=-\frac{\hbar^2}{2m}\frac{\partial^2\Psi}{\partial \rho^2}
+\frac{m\omega_0^2}{2}\rho^2\Psi+gb|\Psi|^2\Psi. 
\eeq
For $g=0$ this is the Schr\"odinger equation of a time-independent 
harmonic oscillator. The evolution of $\psi$ has thus 
been conveniently mapped to the simple solution of an auxiliary stationary system.
Choosing $b(0)=1$, $\dot{b}(0)=0$ 
the ``auxiliary''
(\ref{td2}) 
and physical (\ref{gpe}) oscillators coincide at $t=0$, so any instantaneous eigenstate of $t=0$, with 
vibrational quantum number $n$ and energy $E_n=\hbar\omega_0(n+1/2)$, 
evolves according to a propagating mode determined by Eqs. (\ref{ans},\ref{alpha}) and the solution of the Ermakov equation $b(t)$, 
\beq
\psi(x,t)=b^{-1/2}e^{\frac{im}{2\hbar}\frac{\dot{b}}{b}x^2}
e^{-iE_n\tau(t)/\hbar} \Psi_n(x/b,0).
\eeq
In general this mode will not coincide with the instantaneous eigenstate of the physical Hamiltonian $H(t)=-\frac{\hbar^2}{2m}
\frac{\partial^2}{\partial x^2}+\frac{1}{2}m\omega(t)^2x^2$, unless $b(t)=[\omega_0/\omega(t)]^{1/2}$ and $\dot{b}(t)=0$, up to the global phase factor $e^{-iE_n\tau(t)/\hbar}$. 
This motivated our proposal in \cite{xi2}: it is an inverse method 
in which, given the initial $\omega_0$ and final frequencies $\omega_f=\omega(t_f)$,  the intermediate trajectory $\omega(t)$ is left undetermined at first and   
the boundary conditions
\beqa
b(0)=1,\;\;\dot{b}(0)=0,
\label{4ca}\\
b(t_f)=(\omega_0/\omega_f)^{1/2},\;\;\dot{b}(t_f)=0
\label{4c}
\eeqa
are imposed at initial and final times $t=0,t_f$ (they also imply a vanishing $\ddot{b}$ at these two times to satisfy Eq. (\ref{erma})). $b(t)$ is then interpolated with some functional form, e.g., a polynomial with enough coefficients to satisfy all conditions, and finally $\omega^2(t)$ is calculated from the Ermakov equation (\ref{erma}). This generates, in particular, very fast phase-space conserving cooling processes  where $\omega^2(t)$ takes during some time interval negative values, i.e., the trap becomes an expulsive potential. 

If $g\neq 0$ 
the coefficient of the non-linear term in the auxiliary equation is generally time dependent. 
Thus, imposing $\dot{b}(t_f)=0$ eliminates the phase-factor
$e^{-\alpha(t_f) x^2}$ 
but nothing guarantees that $\Psi(\tau(t_f))$ is proportional to the 
instantaneous eigenstate of the GP equation at $t_f$.   
A way out, in principle, is to make the coupling coefficient time-dependent 
with the aid of a Feshbach  resonance as $g(t)=g_0/b(t)$, with $g_0$ constant.
The resulting auxiliary equation has then time-independent coefficients,
\beq
\label{td3}
i\hbar\frac{\partial \Psi}{\partial \tau}=-\frac{\hbar^2}{2m}\frac{\partial^2\Psi}{\partial \rho^2}
+\frac{m\omega_0^2}{2}\rho^2\Psi+g_0|\Psi|^2\Psi. 
\eeq
and can be solved in the form  $e^{-i\mu\tau(t)/\hbar} \Psi(x/b,0)$,
where $\mu$ is the chemical potential for the initial trap,
so that 
\beq\label{psi1}
\psi(x,t)=b^{-1/2}e^{i\frac{im}{2\hbar}\frac{\dot{b}}{b}x^2} e^{-i\mu\tau(t)/\hbar} \Psi(x/b,0),  
\eeq
and the same inverse method described for the Schr\"odinger equation can now be
applied to design a fast frictionless process for the ground state condensate. 
One can easily check that, keeping $b(t)=b_f$ constant for $t>t_f$, 
which results in $\omega(t)=\omega_f$ and $g=g_0(\omega_f/\omega_0)^{1/2}$
for $t>t_f$, 
the solution $\psi(x,t)$ of (\ref{gpe}) given by (\ref{psi1})
becomes stationary, with a new scaled chemical potential $\mu/b(t_f)^2$.     
%

Other special case is a ``Thomas Fermi'' (TF) limit, keeping $g$ constant.
Using a modified Ermakov equation and a different time scaling  
\beq
\ddot{b}+\omega(t)^2 b=\frac{\omega_0^2}{b^2},\;\;\;\; 
\tau(t)=\int_0^t \frac{dt'}{b},
\label{erma1}
\eeq
render an auxiliary equation with time-independent coefficients 
for the non-linear and harmonic potential terms.  
If $g|\Psi|^2/(\hbar\omega_0)\gg 1$ the kinetic term may  be neglected, 
\beq
i\hbar\frac{\partial \Psi}{\partial \tau}=
\frac{m\omega_0^2}{2}\rho^2\Psi+g|\Psi|^2\Psi. 
\eeq
This equation can also be solved
by separation of variables, $\Psi(x/b,\tau)=e^{-i\mu \tau/\hbar}
\Psi(x/b,0)$, and $\psi(x,t)$ takes again the form of Eq. (\ref{psi1}),  
with different values for $\mu$, $\tau$, $b$, and the initial wavefunction.  
Note that this TF 
approximation is carried out in the auxiliary equation, and not 
at the level of the original GP equation, 
since that would imply a frozen density \cite{Cas,Shlyap96}.    
From the modified Ermakov equation in (\ref{erma1}), the inversion method to find a frictionless trajectory $\omega(t)$ requires 
in this 1D-TF scenario to change the boundary condition at $t_f$ in (\ref{4c}) to   
$b(t_f)=(\omega_0/\omega_f)^{2/3}$, with $\ddot{b}(0)=\ddot{b}(t_f)=0$ as before.   
\section{Two and three dimensional traps}
The manipulations in 1D suggest for 2D and 3D a wavefunction ansatz of the 
form \cite{Shlyap96} 
\beq
\label{ansa}
\psi({\bf{r}},t)
=b^{-d/2}e^{\frac{im r^2}{2\hbar}\frac{\dot{b}}{b}}\,\phi({\bf{r}},t),
\eeq
where $d$ is the dimension, $r=(x^2+y^2)^{1/2}$ in 2D or $r=(x^2+y^2+z^2)^{1/2}$ in 3D.  
This form guarantees an auxiliary equation without first spatial derivatives.
  
With $\rhov={\bf{r}}/b$ and a notation for the wavefunctions parallel to the 1D case there results,
by substituting (\ref{ansa}) into the 2D or 3D GP equations,  
\beq
i\hbar \frac{\partial \Psi}{\partial \tau}\left(\frac{d\tau}{dt} b^2\right)
=
-\frac{\hbar^2}{2m}\Delta_\rho \Psi+\frac{m}{2}
\left[\omega^2(t)+\frac{\ddot{b}}{b}\right]\rho^2b^4\Psi
+\frac{g}{b^{d-2}}
|\Psi|^2\Psi,
\eeq
where the Laplacian should be adapted to the dimension,
$\Psi=\Psi(\rhov,\tau)$, and $\tau$ has not been specified yet.
(This equation includes the case $d=1$ too 
by substituting $r\to x$ and the Laplacian by a second derivative.) 

In 2D, the ordinary Ermakov equation (\ref{erma}) and the $\tau$  
in Eq. (\ref{tau2}) are the optimal choice since all coefficients in the auxiliary equation (assuming a constant $g$) become time independent, even outside the Thomas-Fermi regime, 
\beq
i\hbar\frac{\partial \Psi}{\partial \tau}=-\frac{\hbar^2}{2m}\Delta_\rho\Psi
+\frac{m\omega_0^2}{2}\rho^2\Psi+g|\Psi|^2\Psi. 
\eeq
This is then the ideal situation for designing a frictionless process by 
shaping $b$ and $\omega$ exactly as in the 1D Schr\"odinger equation, i.e., using 
(\ref{4ca}) and (\ref{4c}).     

Finally, the case $d=3$ is considered. It is somewhat similar to 1D, in the sense  
that the generic case leads to time-dependent coefficients
in the auxiliary equation. Similarly to 1D, by using Eqs. (\ref{erma}) and (\ref{tau2}) the time-independence of the coefficients in the auxiliary equation 
would require now a time dependent coupling of the form $g(t)=g_0 b(t)$; alternatively, 
in the Thomas-Fermi regime and with $g$ constant, all coefficients become 
time independent 
with 
\beq
\ddot{b}+\omega(t)^2 b=\frac{\omega_0^2}{b^4},\;\;\;\; 
\tau(t)=\int_0^t \frac{dt'}{b^3},   
\eeq
and in this case the boundary condition for $b(t_f)$ in (\ref{4c}) 
should be modified to $b(t_f)=(\omega_0/\omega_f)^{2/5}$, assuming  again $\ddot{b}(0)=\ddot{b}(t_f)=0$.   
\section{Examples}
%
\begin{figure}[t]
\begin{center}
%
\includegraphics[width=0.40\linewidth]{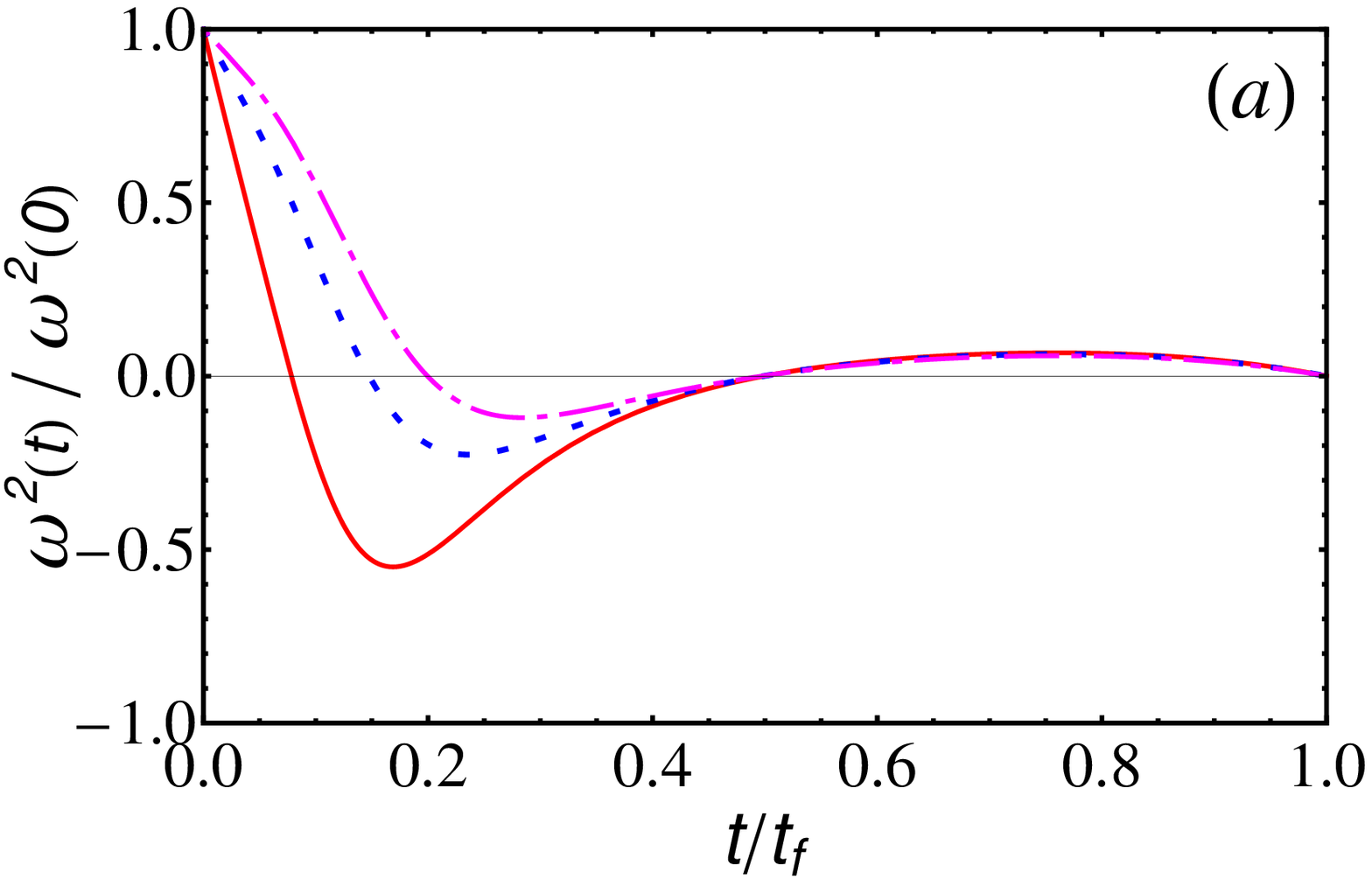}
\includegraphics[width=0.40\linewidth]{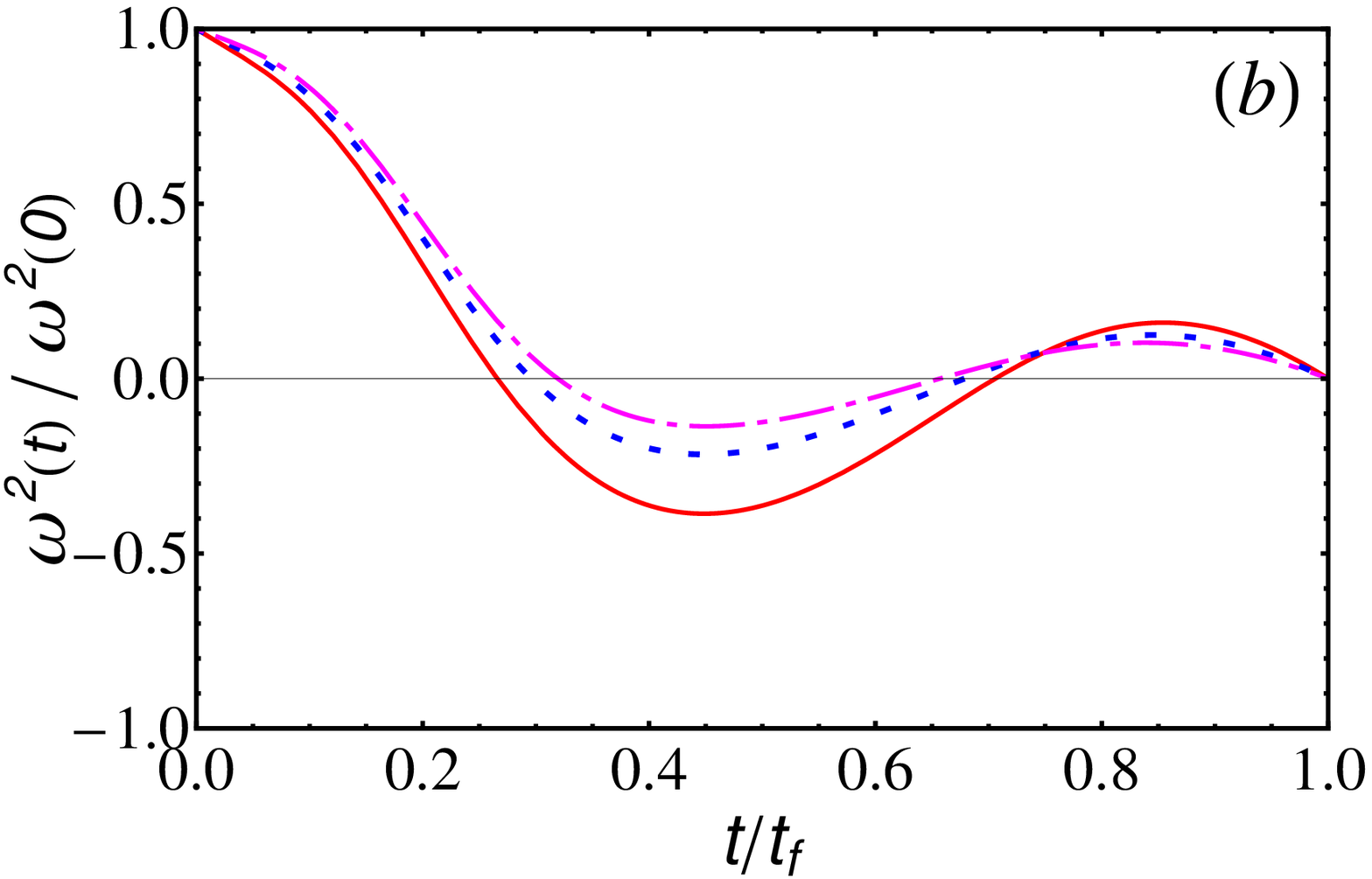}
\end{center}
\caption{\label{fig1}(color online). The squared frequency $\omega^2(t)$ for 
an expansion from $\omega_0=250\times2\pi$ Hz to $\omega_f=2.5\times2\pi$ Hz 
in $t_f=6$ ms (a) Polynomial form $b=\sum_{j=0}^5 a_j t^j$; (b) Exponential of a polynomial
$b=\exp{\sum_{j=0}^5 c_j t^j}$. In both figures: 1D, TF (solid, red line); 2D, or 1D with $g(t)=g_0/b(t)$, or 
3D with $g(t)=g_0b(t)$ (dotted, blue line); 3D, TF (dot-dashed, magenta line).}
\end{figure}
\begin{figure}[h]
\begin{center}
%
\includegraphics[width=0.38\linewidth]{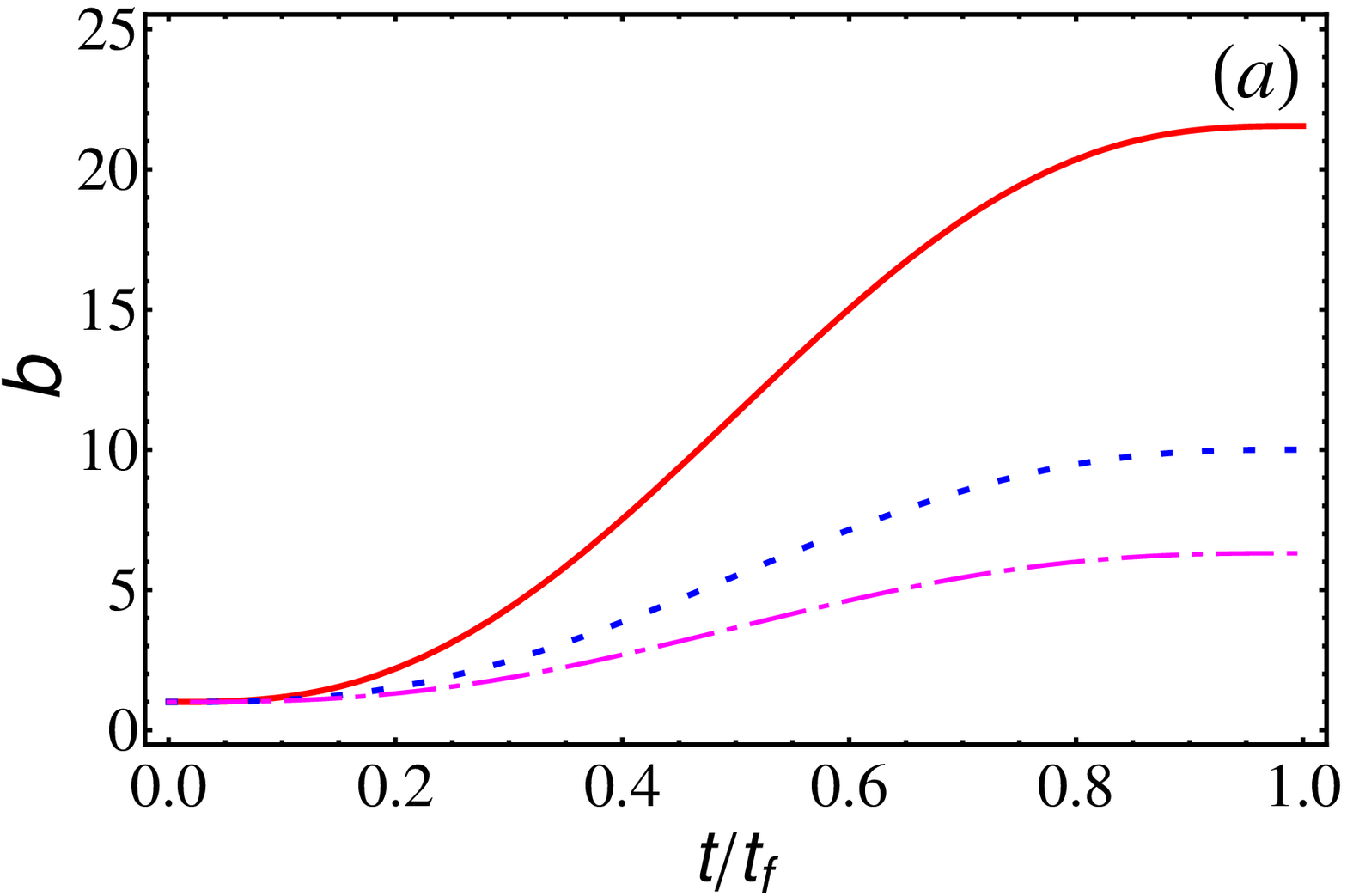}
\includegraphics[width=0.38\linewidth]{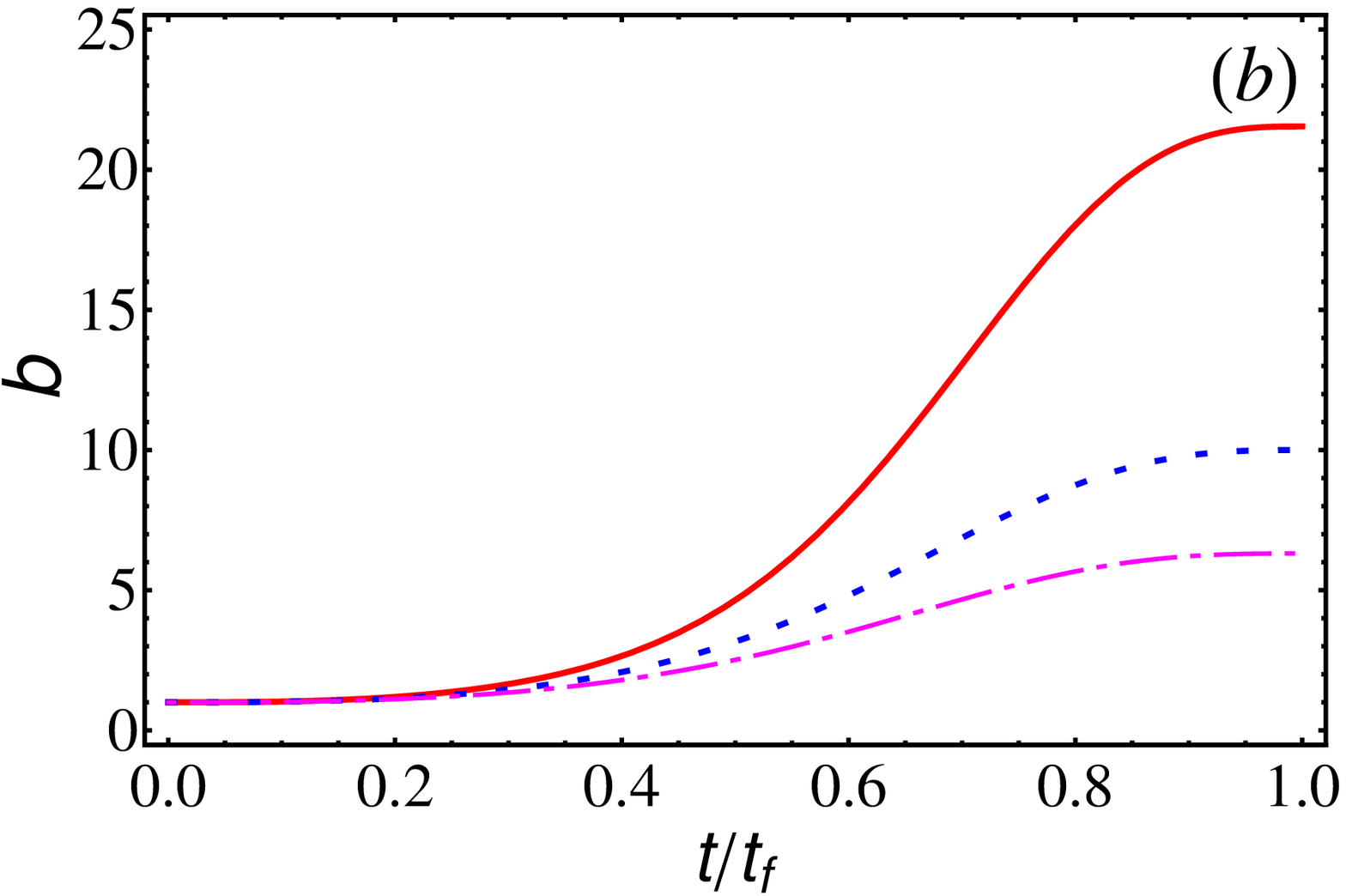}
\end{center}
\caption{\label{fig2}(color online). $b$ corresponding to Fig. 1, $b(t_f)=(\omega_0/\omega_f)^{2/\nu}$, with $\nu=3$ (solid, red), 
$\nu=4$ (dotted, blue), and $\nu=5$ (dot-dashed, magenta). (a) Polynomial $b$, 
(b) $b$ is the exponential of a polynomial.}  
\end{figure}
Let us consider an expansion reducing the frequency 
100 times from $250\times2\pi$ Hz to $2.50\times2\pi$ Hz in 6 ms. This time is too short for the condensate to follow
any $\omega(t)$ adiabatically at all times \cite{xi2} but with our designed trajectories 
and thanks to the expulsive interval which accelerates the spreading, 
the final state would indeed be the same, up to a global phase, than 
the state obtained if a slow process could be implemented for such an expansion.
For the regular Ermakov equation (\ref{erma}), a polynomial $b(t)$, 
and $\omega_0\gg\omega_f$, a simple estimate is that a repulsive time interval is 
necessary for $t_f< 1/(2\omega_f)$.    
Figure 1 shows frequency trajectories for the different cases discussed 
before and two different ansatz of $b$. Higher powers of $b$ in the 
right hand side of the Ermakov-like equations (corresponding to higher dimensions in the TF regime) imply a smaller increment for $b$ during the expansion, 
see Fig. 2,  which makes the change of $\omega$ smoother as well. 
It is remarkable that for a fixed $g$ (for 2D, or the TF regimes in 1D and 3D), 
the non-linearity does not play any role in the 
design of optimal (frictionless) frequency trajectories. They only depend on the initial and final frequencies, the available time $t_f$ and the functional form chosen for $b(t)$.
\section{Discussion}
%
%
%
%
%
%
%
We will provide here some complementary details and relate the results to other works.   
An important remark on the TF approximation used for 1D and 3D geometries 
is that
the non-linear coupling cannot be arbitrarily strong. 
The condition $g|\Psi|^2/(\hbar\omega_0)\gg1$
should be compatible with the derivation of
the 1D GP equation \cite{Salas} in a weak interaction limit, i.e., $a_s  |\psi|^2<<1$, 
where 
$a_s$ is the 
$s$-wave scattering length.   
              
For completeness we should mention an alternative to the steps
given after Eq. (\ref{alpha}). 
We may as well impose that the coefficient multiplying $\rho^2 b^4\Psi$ 
must vanish instead of becoming a non-zero constant \cite{Wu07,Kumar08,Kh09}. 
This amounts to imposing
%
$\ddot{b}+\omega(t)^2 b=0$
%
instead of the Ermakov equation (\ref{erma}). Proceeding as in Sec. \ref{odt}
with $\tau$ given by Eq. (\ref{tau2}), the resulting auxiliary equation becomes 
\beq
i\hbar\frac{\partial \Psi}{\partial \tau}=-\frac{\hbar^2}{2m}\frac{\partial^2\Psi}{\partial \rho^2}
+gb|\Psi|^2\Psi,  
\label{alt}
\eeq
which is not an equation for the harmonic oscillator 
but for a condensate without confining external fields and with a, generically, 
time dependent 
non-linear coupling factor. Adapting the time dependence of $g$ as $g(t)=g_0/b(t)$, 
this method provides, from known analytical solutions of 
Eq. (\ref{alt}) with a constant factor $g(t)b(t)=g_0$, 
explicit solutions that have been used in the context 
of soliton dynamics \cite{Wu07,Kumar08,Kh09}. 
While the solutions $\psi(x,t)$ for the same $\omega(t)$ and initial conditions should of course be equivalent to the ones obtained with the ordinary Ermakov equation, we find the later better suited for the application of our inverse technique. 

 
 
In summary, it is possible to take a Bose-Einstein condensate in a very short 
time from an initial harmonic trap to a final one without excitations, 
by choosing the time dependence of the frequency according to the Ermakov equation or its modifications after matching the time dependence of a scaling factor 
to suitable boundary conditions. In 1D and 3D traps this requires either a simultaneous change of the time-dependence of the coupling, or a Thomas-Fermi type of regime. 2D traps are privileged in this respect and do not require either of these conditions. Their 
peculiar symmetry properties were already noticed by Pitaevskii and Rosch \cite{PR}.  
Indeed, the 2D geometry allows for an extension of the present results beyond the GP equation 
framework by expanding perturbatively the field operator
around the condensate wavefunction, and treating the perturbation with an ansatz 
parallel to (\ref{ansa}) and the same phase \cite{Shlyap96}.

\ack{We acknowledge funding by Projects No. GIU07/40, FIS2006-10268-C03-01,
60806041, 08QA14030, 2007CG52, S30105, ANR-09-BLAN-0134-01, R\'egion 
Midi-Pyr\'{e}n\'{e}es, Juan de la Cierva Program,
and the German Research Foundation (DFG).}


\section*{References}
\begin{thebibliography}{99}
%
%
\bibitem{Ronnie} Rezek Y and Kosloff R 2006 {\it N. J. Phys} \textbf{8} 83
\bibitem{Shlyap96} Kagan Y, Surlov E L and Shlyapnikov G V 1996 
{\it Phys. Rev. A} \textbf{54} R1753  
\bibitem{cd} Castin Y and Dum R, Phys. Rev. Lett. 1997 \textbf{79} 3553  
\bibitem{Uhlmann} Uhlmann M 2009 {\it Phys. Rev. A} \textbf{79} 033601 
\bibitem{xi2} Chen X, Ruschhaupt A, {Schmidt S}, {del Campo A},
Gu\'ery-Odelin D and Muga J G 2009 arXiv:0910.0709
\bibitem{Schmidt} Schmidt S, Muga J G and Ruschhaupt A 2009
{\it Phys. Rev. A} \textbf{80} 023406
\bibitem{root2} Chen Xi, Muga J G, del Campo A and Ruschhaupt A 
2009 arXiv:0908.1483 
\bibitem{LR69} Lewis H R and Riesenfeld W B 1969 
{\it J. Math. Phys.} \textbf{10} 1458   
\bibitem{Palao} Palao J P, Muga J G and Sala R 1998 {\it Phys. Rev. Lett.} \textbf{80}  5469
\bibitem{Salas} Salasnich L, Parola A, and Reatto L 2002 {\it Phys. Rev. A}
\textbf{65} 043614
\bibitem{invGP} Bassalo J M F, Alencar P T S, Silva D G, Nassar A B and  Cattani M
2009 eprint arXiv:0902.3125
\bibitem{Cas} Castin Y and Dum R 1996 {\it Phys. Rev. Lett.} \textbf{77}
5315  
\bibitem{Wu07} Wu L, Zhang J F and Li L 2007 {\it N. J. Phys} \textbf{9} 69 
\bibitem{Kumar08} Kumar V R, Radha R and Panigrahi P K 2008 {\it Phys. Rev. A} \textbf{77} 023611 
\bibitem{Kh09} Al Khawaja U 2009 {\it J. Phys. A: Math. Theor.} \textbf{42} 265206
\bibitem{PR} Pitaevskii L P and Rosch A, Phys. Rev. A 1997 \textbf{55} R853
\end{thebibliography}
\end{document}